# Enhanced pseudocapacitance from finely ordered pristine $\alpha$-MnO$_2$ nanorods at favourably high current density using redox additive

Niraj Kumar[ab], K. Guru Prasad[ab], Arijit Sen[ab*], and T. Maiyalagan[a]

[a]SRM Research Institute, SRM University, Kattankulathur-603203, India

[b]Department of Physics and Nanotechnology, SRM University, Kattankulathur-603203, India

*Corresponding Author E-mail: *arijit.s@res.srmuniv.ac.in*

**Abstract**

A flexible technique is developed using hydrochloric acid to modify the redox reaction between potassium permanganate and sodium nitrite in order to grow ultrafine $\alpha$-MnO$_2$ nanorods, hydrothermally. The nanorods grown were 10-40 nm diameters in range. Not any crack, fissure, imperfection or dislocation is observed in the nanorods suggesting it to be finely ordered. Structure, phase and purity of as developed nanorods were determined using X-ray diffraction (XRD), Fourier transform infrared spectroscopy (FTIR) and Energy-dispersive X-ray spectroscopy. Peseudocapacitance of $\alpha$-MnO$_2$ nanorods was tested using a three electrode system. Considerably very high pseudocapacitance value of 643.5 F/g at 15 A/g current density was calculated from the galvanostatic discharge current measurement. Also excellent cyclability is observed with high retention of 90.5% after 4000 cycles. Highly uniform and confined morphology of the nanorods helps smooth the electron dynamics between electrode/electrolyte interfaces resulting in superior performance. Most importantly, the use of potassium ferricyanide as redox additive to KOH electrolyte was proved to be quite effective as it provides extra redox couple [Fe(CN)$_6$]$^{3-}$/[Fe(CN)$_6$]$^{4-}$ which helps in further smoothening of electron transition thereby resulting in considerably superior pseudocapacitive performance.

*Keywords*: Hydrothermal; nanorods; ultrafine; pseudocapacitor; redox additive; galvanostatic

# 1. Introduction

The developments in modern lifestyles have resulted in ever increasing demands for energy sources which have proved a great threat for the natural resources of fossil fuels. To tackle this crisis, there is a need to harvest sustainable energy plus develop new technologies for improved energy storage and conversion [1-3]. Supercapacitors are one such solution due for their high power density and long cycle life with quick charge/discharge rate compared to batteries [4-9]. In contrast to conventional electrostatic capacitors, they provide high energy density [4, 5]. Generally, supercapacitors found applications where large power is needed for short interval like breaking systems of electric and hybrid vehicles; emergency power shutdown and power stabilizers in low power devices [10,11]. Supercapacitors are classified into electric double layer capacitors (EDLC) where the charges are stored electrostatically through separation of charges and pseudocapacitors (PC) where the charges are stored faradaically involving electrochemical redox reaction, electrosorption and intercalation of ions or atoms between electrode and electrolyte [12, 13]. Faradaic storage mechanism of charge makes pesudocapacitors meritorious by giving higher energy density than EDLC [14-16]. Transition metal oxide is considered as an ideal pesudocapacitor electrode material attributing to their intrinsically high pseudocapacitive behaviour of fast and reversible surface redox reactions [17-19].

Till date various transition metal oxides namely manganese dioxide ($MnO_2$) [20-22], cobalt oxide ($Co_3O_4$) [8, 9, 23], nickel oxide (NiO) [24-26], and ruthenium dioxide ($RuO_2$) [27] as electrodes of pseudocapacitors have been studied to achieve superior specific capacitance values. Among them $RuO_2$ is considered to be the most effective electrode material, but anyhow due for its low availability and high cost its implementation is often discouraged [27]. As an alternative, $MnO_2$ can be a promising electrode material owing to its high theoretical capacitance (1370 F/g), low cost and environmental benignity [28-32]. Moreover in contrast to NiO and $Co_3O_4$, it offers larger working potential range around 1V, which is favourable for getting higher energy density [10]. In addition, the tunnelled crystal structures of $MnO_2$ can favour ion diffusion better than other spinel structures and thus can facilitate redox reactions [33].

$MnO_2$ exits in different crystallographic polymorphs, $\alpha$-, $\beta$-, $\gamma$ -, $\delta$- and $\varepsilon$-$MnO_2$ showcasing its diversified structural nature [34, 35]. $\alpha$- $MnO_2$ is believed to perform better than its other polymorphs counterpart due for its large 2×2 tunnel structure with approximately 0.46 nm size and high surface area [36, 37]. Fabrication of $MnO_2$ with fast and reversible surface redox reactions is a challenge which can be counter through shortening of diffusion length between electrode and electrolyte interface [38]. Synthesis of $MnO_2$ nanomaterials can be a solution to this, as sharp edges and extremely low dimensions exhibited by nanomaterials can considerably reduce the diffusion length when compared with their bulk counterparts [38]. $α$- $MnO_2$ in shapes of octahedral molecular sieves [39], nanosheets [40], nanoflowers [41], hollow spheres [42], nanobelts [43], nanowires [34] and nanorods [44-46] have been prepared in recent times. Zhaoxia et al. [47] have studied the capacitive performance of $α$- $MnO_2$ naorods in 1M KOH alkaline aqueous electrolyte solution. Xiaohui et al. [36] have tested branched $α$- $MnO_2$ naorods for supercapacitors applications in 1M $Na_2SO_4$ neutral solution.

However there is still one more prospect to increase the cycle life and transfer rate for redox reactions at higher current densities of pseudocapacitors with introduction of redox additives (potassium ferricyanide, $K_3Fe(CN)_6$ to electrolyte (KOH). Cuimei et al. [48] have demonstrated ultrahigh specific capacitance for $Co(OH)_2$ electrode using and 0.08 M $K_3Fe(CN)_6$ as redox additives in 1 M KOH as electrolyte. Sandipan et al. [17] have reported an increment of about 7 times in energy density with the introduction of $K_3Fe(CN)_6$ as redox additive to KOH electrolyte for interconnected network of $MnO_2$ nanowires as electrode.

Herein we report a simple hydrothermal method for synthesis of finely ordered and ultrapure $α$- $MnO_2$ nanorods as analyzed from the XRD, FTIR, EDS, FESEM and HRTEM characterizations. Pseudocapacitive performance of pristine $α$- $MnO_2$ nanorods are studied using $K_3Fe(CN)_6$ as redox additive to KOH electrolyte. Cyclic voltammetry and galvanostatic charge/discharge measurements were undertaken by three electrode system. An interestingly ultrahigh pseudocapacitive performance and capacity retention is acknowledged due mainly for redox additive and ultra finesse of nanorods.

## 2. Experimental

### 2.1 Materials

Potassium permanganate ($KMnO_4$), Sodium nitrite ($NaNO_2$), Hydrochloric acid (HCl), Potassium hydroxide (KOH) and Potassium ferricyanide ($K_3Fe(CN)_6$) were purchased from Sigma Aldrich. For preparing solutions, de-ionized water was used.

## 2.2 Synthesis

In a typical synthesis of α-MnO$_2$ nanorods, KMnO$_4$ and NaNO$_2$ were mixed using magnetic stirring in 2:3 molar ratios to form a homogeneous solution of 39 ml. Then 1 ml solution of 0.3 M HCl was prepared which was added slowly and steadily into the solution under vigorous stirring. The final solution was exposed to hydrothermal pressure by placing it in autoclave of 80% total capacity and maintaining temperature of 170$^o$C for 12 h. After the hydrothermal process the residual liquid was discarded and precipitate was collected. The precipitate was washed thoroughly and calcined at 300$^o$C for 3 h to obtain α-MnO$_2$ nanorods with high purity.

## 2.3 Electrode preparation

The electrode was made by grinding LiMn$_2$O$_4$, super P carbon and PVDF together as active material, conductor and binder in weight ratio of 8:1:1, respectively. After 1 h grinding, slurry was formed by use of adequate amount of N-Methyl-2-pyrrolidone. 0.7 mg of as prepared material was then evenly coated on stainless steel plate (304 grade SS plate with 0.3 mm thickness) covering 1cm$^2$ area which acts as a current collector. The coated plate was kept in Hot Air Oven for drying at 100˚C for 6h. Similar procedures were followed with MnO$_2$ nanorods as active material.

## 2.4 Characterization

PAN analytical X' Pert Pro diffractometer was used for X-ray crystallography at 1.5406 Å Cu-Kα rays of wavelength, 30 mA tube current, 40 kV and 10-80 degree 2 θ range. Fourier transform infrared spectroscopy was carried out on with Perkin Elmer Spectrophotometer at 400-4000 cm$^{-1}$ wavenumber using KBr pellet method. Physical dimensions of as prepared sample were characterized using 'Quanta 200 FEG FE-SEM', Field emission scanning electron microscope and 'HR-TEM, JEM-2010, 200kV', High resolution Transmission electron microscope. Energy dispersive X-ray microanalysis (EDXMA or EDS) was carried on with 'Bruker 129 ev' with Espirit software. Cyclic voltammetry and galvanostatic charge/discharge measurement were taken with Biologic SP300 using three electrode system. The as synthesized material served as working electrode against platinum wire and silver/silver chloride (Ag/AgCl) in 3M KCl as counter and reference electrodes, respectively.

# 3 Results and discussions

## 3.1 Structural analysis

XRD pattern of sample prepared with $KMnO_4$, $NaNO_2$ and HCl as precursors is presented in fig. 1a. The prime diffraction peaks at $2\theta$ = 12.7, 18.1, 28.8, 37.4, 49.8, 60.2 corresponds to tetragonal structure of $\alpha$-$MnO_2$ as prescribed in JCPDS 44-0141 card number. High pristine nature of as prepared samples can be confirmed from the absence of any impurity peaks. Figure 1b represents the FTIR spectrum of $\alpha$-$MnO_2$. Mn-O vibrations in $MnO_6$ octahedra can be acknowledged through bands observed at 716, 527 and 437 $cm^{-1}$ [34]. Absence of impurity peaks further confirms for pristine nature of as prepared $\alpha$-$MnO_2$. Energy-dispersive X-ray spectroscopy measurements are visible in fig. 1c. Sharp peaks for manganese and oxygen elements with no traces of other impurity peaks can be an evidence for pristine nature of as prepared $\alpha$-$MnO_2$.

## 3.2 Morphological analysis

Physical appearance of as prepared $MnO_2$ sample can be visualized from fig 2. FESEM images shown in fig. 2(a-d) confirm the 1D morphology for $MnO_2$ with diameters in range of 10-40 nm. The as prepared nanorods can be seen to be finely ordered as there are no dislocations or defects. Ultra sharp one dimensional growth with no cracks or fissures is observed for as synthesized $\alpha$-$MnO_2$ nanords. The growth of these ultrafine $MnO_2$ nanorods is further validated by HRTEM analysis. HRTEM image shown in fig. 2(c,d) confirms for the ultrafine nature. *d*-sapcing of 0.314 nm in growth direction (310) as shown in inset fig.2d is in agreement with the JCPDS 44-0141 data and is an evidence for formation of $\alpha$-$MnO_2$.

## 3.3 Growth Kinetics

The chemical kinetics involving synthesis of $\alpha$-$MnO_2$ nanorods primarily revolves around the redox reaction as described below [49]:

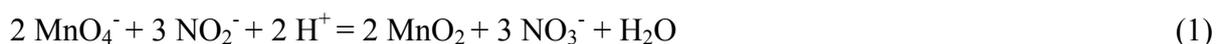

$$2\ MnO_4^- + 3\ NO_2^- + 2\ H^+ = 2\ MnO_2 + 3\ NO_3^- + H_2O \qquad (1)$$

For feasible reaction optimizations in precursors contents are mandatory. The stoichiometric molar ratio of $MnO_4^-$ and $NO_2^-$ should be 2:3 to yield $MnO_2$. This redox reaction involves nitrite ions ($NO_2^-$) as reducer and permanganate ions ($MnO_4^-$) as oxidizer. The necessary protons ($H^+$ ions) required as per Le Chatelier principle comes from hydrochloric acid (HCl). However, principle optimization comes from the addition of HCl into reaction medium. If inadequate amount of HCl is supplied during reaction, then there are chances for formation of $MnCl_2$ due for creation of $Mn^{2+}$ ions as per below mentioned equation:

$$2\ KMnO_4 + 5\ NaNO_2 + 6\ HCl = 2\ KCl + 2\ MnCl_2 + 5\ NaNO_3 + 3\ H_2O \qquad (2)$$

Therefore, to prohibit the above reaction (eq. 2), HCl should be added meticulously which can be possible through dropwise addition. In this way, the reaction medium can be enriched with $Mn^{4+}$ ions which can favour $MnO_2$ formation as $MnO_2$ crystal structure comprises $MnO_6$ octahedra units and in each unit six oxygen atoms surrounds single $Mn^{4+}$ ion forming interlinked tunneled structures [35, 50]. Moreover, optimal acidic pressure works in tandem with hydrothermal pressure in slimming down the morphologies, resulting in formation of ultrafine α-$MnO_2$ nanords. Still new explorations are needed to be done in near future.

## 3.2 Electrochemical studies

A three electrode system was designed to carry out the electrochemical studies. As prepared electrode with α-$MnO_2$ nanorods works as working electrode whereas platinum foil and Ag/AgCl (dipped in 3M KCl), works as counter and reference electrodes, respectively. The test was carried out in aqueous electrolyte solution of 3 M KOH + 0.1 M $K_3Fe(CN)_6$. Figure 3a shows the cyclic voltammetry (CV) measurements of α-$MnO_2$ nanords at varying scan rates of 5, 10, 15, 20, 25 and 30mV/s in voltage range of 0.6 to -0.4V. Sharp redox peaks are conspicuous. This gives an indication for promising pseudocapacitance performance of as prepared electrode. Sharp anodic oxidation and cathodic reduction peaks are observed at 0.47 V and 0.27 V, respectively. Redox peaks are prominent at 5 mV/s which keeps on decreasing with increasing CV curve area at higher scan rates. The sharp redox kinetics observed can be assigned to two simultaneous phenomena, first being an effect from KOH electrolyte and second from $K_3Fe(CN)_6$:

(1) Redox kinetics involving KOH electrolyte is further classified into two simultaneous phenomena:

(a) Non-faradaic activity involving adsorption or desorption of $K^+$ ions by $MnO_2$ electrode surface [51, 52]:

$$(MnO_2) + K^+ + e^- \leftrightarrow (MnO_2^-\ K^+)_{surface} \qquad (3)$$

(b) Faradaic activity involving intercalation or de-intercalation of $K^+$ ions by $MnO_2$ interstitial site:

$$MnO_2 + K^+ + e^- \leftrightarrow MnO.OK \quad (4)$$

(2) Redox kinetics involving $K_3Fe(CN)_6$ [53]:

$$K_3Fe(CN)_6 + e^- \leftrightarrow K_4Fe(CN)_6 \quad (5)$$

Each of the above processes is concurrent and fully reversible requiring a charge. As capacitive performance mainly depends upon storage of charge, so more number of charge involvement in reversible chemical kinetics relates to improvement in energy storage property. In electric double layer capacitors, some amount of charge is acquired at surface by adsorption (eq. 3) and helps in charge storage. In pseudocapacitors, charge storage is increased by involvement of extra amount of charge through Faradaic reaction as some charge are taken up electrochemically by interstitial site of electrode (eq. 4) through conversion of $Mn(IV) \leftrightarrow Mn(III)$ and adds up for total amount of charge acquisition. As all the kinetics are fully reversible, so the charge gained during reduction is accompanied by loosening of charge during oxidation forming a redox couple. This redox couple is supposedly enhanced by the addition of $[Fe(CN)_6]^{3-}/[Fe(CN)_6]^{4-}$ redox couple (eq. 5). Normally, during charging hexacyanoferrate Fe(III) is reduced to Fe(II) and being reversible reaction, further provide electrons during oxidation which helps in smoothening of charge acquisition and thus boost the pseudocapcitor performance. Superior capacitive performance of pseudocapacitors involves hybrid mechanism between battery and electric double layer capacitance which is quite complex and still needs further studies. Anyhow our assumptions are validated by the experimental studies carried out using galvanostatic charge/discharge measurements in the following section.

Also it can be seen that with increasing scan rates there is drastic increase in the curve area suggesting for higher capacitive performance. However, exact pseudocapacitance is not calculated using CVs due for the changing discharge current across the potential range [17]. For this galvanostatic charge/discharge studies were carried out. Figure 4b shows the charge/discharge profiles in voltage range from -0.4 to 0.6 V versus time at different current densities of 15, 20, 25 and 30 A/g. The discharge time decreases with increasing current. High current densities were incorporated to make sure that discharge time should be less. Normally, the supercapacitors must possess low discharge time than conventional Li-ion batteries for their unique applications needing instant energy supplies. This can be achieved at high current density, but unfortunately they lose their performance at this point due for quasi reversible redox reactions [17]. So most of pseudocapacitors are reported at low current densities of 0.2 - 5 A/g [6, 10, 17] for $MnO_2$ based electrodes. However, we have made an effort to lower the discharge time without much sacrificing the pseudocapacitive performance. The pseudocapacitance (C in F/g) was calculated using following equation [10]:

$$C = I \Delta t / m \Delta V \tag{6}$$

Where, $I$, $\Delta t$, $m$ and $\Delta V$ represents discharge current in ampere (A), discharge time in seconds (s), mass of the active material in grams (g) and potential difference in volts (V), respectively. Pseudocapacitance values of 643.5, 270 and 185 F/g at current densities of 15, 20 and 25 A/g are achieved. The presented values are much superior than most of the previous reports such as 164 F/g at 16 A/g for branched $MnO_2$ nanorods [36], 106 F/g at 0.5 A/g for $MnO_2$/CNT [10] and 96 F/g at 20 A/g for $MnO_2$ thin film [54]. Fortunately, this large difference was achieved mainly due for ultrafine and uniform nature of as prepared nanorods as visualized from FESEM imges (fig. 2). The nanorods are visible to be finely ordered even at higher magnifications (fig. 2b) and so could have offered very low charge diffusion path and also their sharp surface edges could have helped in smoothening the electron transition between the electrode/electrolyte interfaces.

Capacity retention values for up to 4000 continuous charge/discharge cycles at 25 A/g is depicted in fig. 3c. It is captivating to observe increment in capacity of about 110% after continuous cycling nearly between 500-1000 cycles. This may be due to increment in reversible redox conversion of Fe(II) to Fe(III). Anyhow after 1000 cycles the capacity came down nearly to its original value and highly stabilized performance was achieved as tested for up to 4000 cycles. Excellent retention capacity of 90.5% is observed after 4000 cycles. This is much higher than the reported 72.5% for $MnO_2$ nanorods [36]. After prolong cycling, the redox assisted conversion of Fe(II) to Fe(III) is supposed to reach equilibrium condition with provision of equal number of Fe(II) and Fe(III) ions [55]. This provides an excellent electron transition pathway thereby stabilizing the performance. Figure 3d showcases constant discharge/discharge cycles for up to 20 cycles. Figure 4(a-d) highlights the stable charge/discharge profiles after 1000, 2000, 3000 and 4000 cycles, respectively.

## 4 Conclusions

Hydrochloric acid content was successfully optimized with respect to hydrothermal pressure to give rise to finely ordered α-$MnO_2$ nanorods. The as prepared nanorods turned out to be pristine in nature as no impurity or other phases were observed being validated from XRD, FTIR and EDS analyses. These nanorods were free from any crack, fissure or dislocation. The uniformity observed in nanorods helped to exhibit promisingly superior pseudocapacitive behaviour. In addition, the use of potassium ferricyanide as redox additive was proved to be very effective. $[Fe(CN)_6]^{3-}/[Fe(CN)_6]^{4-}$ redox couple facilitated the electron dynamics between electrode and electrolyte which eventually paved the way for attaining much higher pseudocapacitance for $MnO_2$ at higher current densities along with stable cyclic behaviour.

**Acknowledgment**

This work was supported by DST Nano Mission, Govt. of India, via Project No. SR/NM/NS-1062/2012.

# Figure captions:

**Figure 1:** (a) XRD, (b) FTIR and (c) EDS patterns for sample prepared with $KMnO_4$, $NaNO_2$ and HCl as precursors using hydrothermal method.

**Figure 2:** (a, b) FESEM and (c, d) HRTEM images of α-$MnO_2$ nanorods with *d*-spacing in (310) plane direction as inset (d).

**Figure 3:** (a) Cyclic voltammetry, (b) Galvanostatic charge/discharge, (c) Capacity retention versus cycle at 25 A/g and (d) First 20 cycles charge/discharge profiles at 25 A/g.

**Figure 4:** Galvanostatic charge/discharge profiles for (a) 1000, (b) 2000, (c) 3000 and (d) 4000 cycles at 25 A/g.

**Figures**

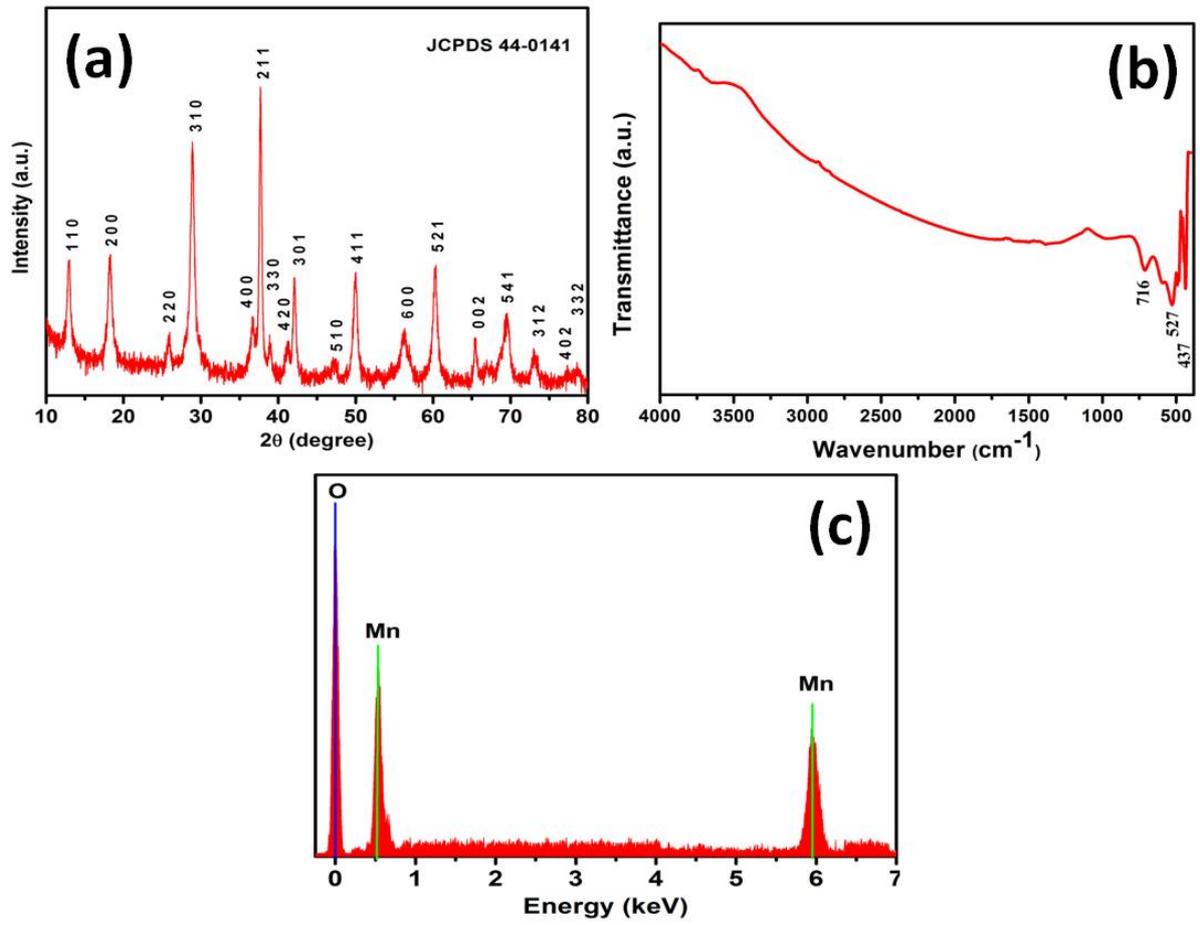

**Figure 1**

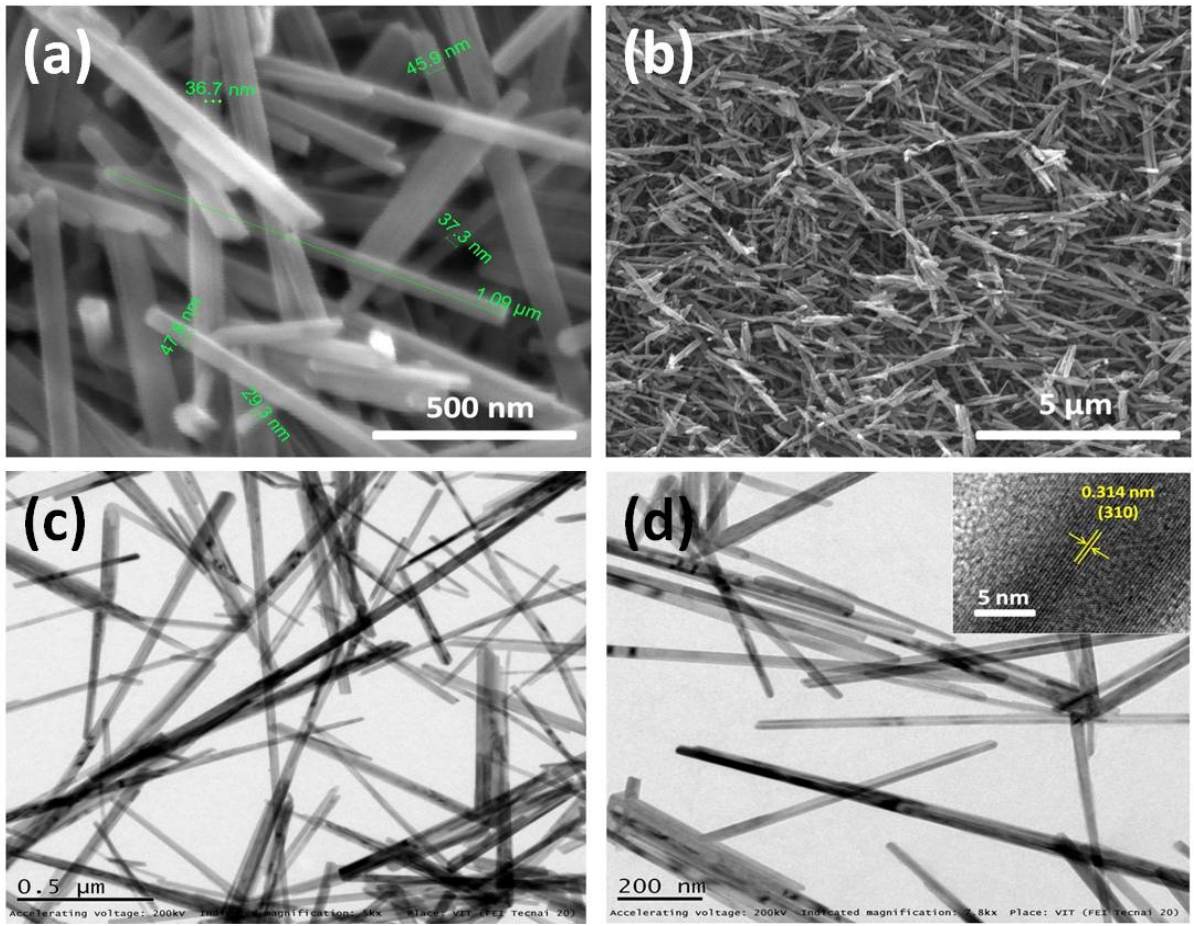

**Figure 2**

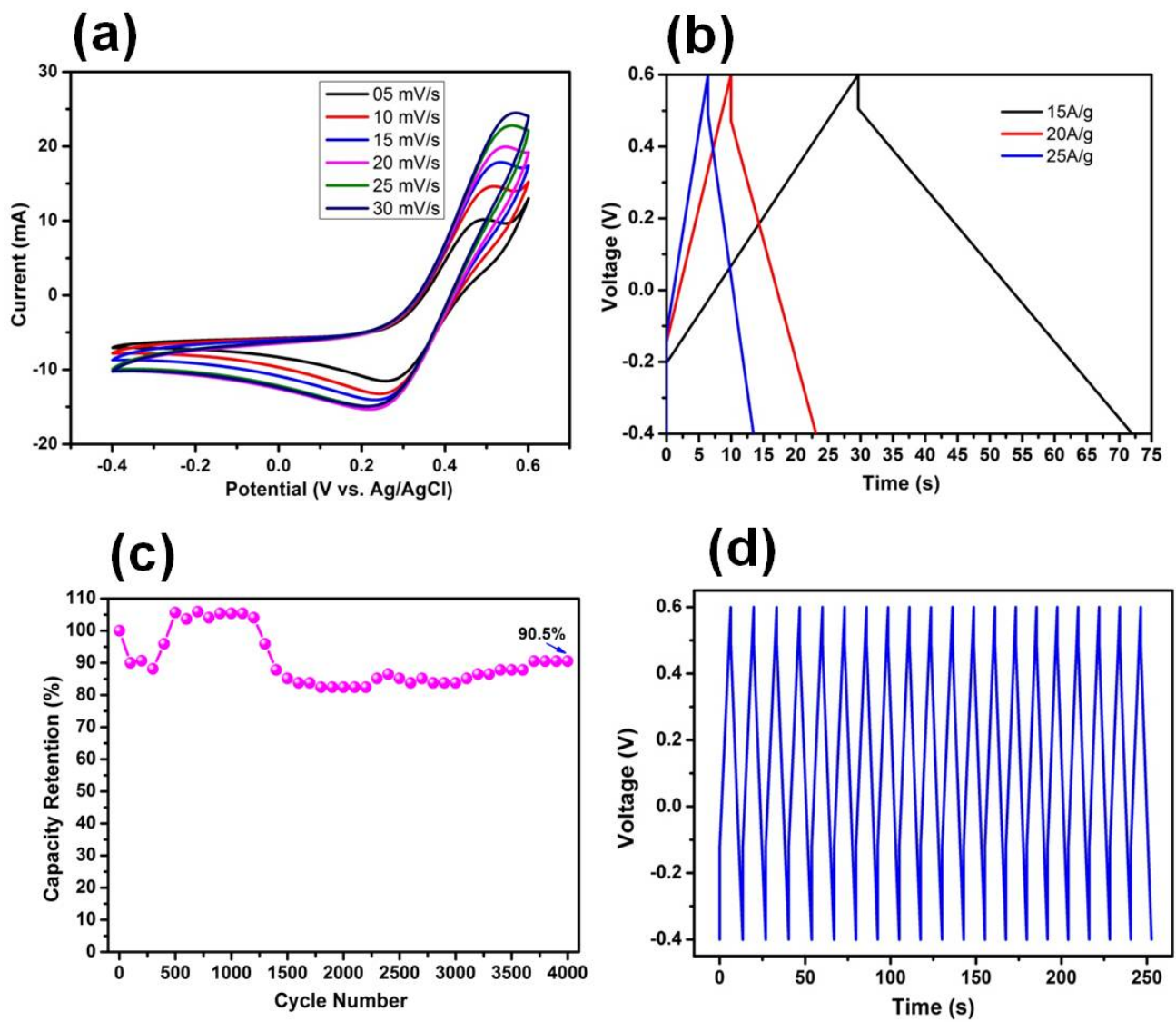

**Figure 3**

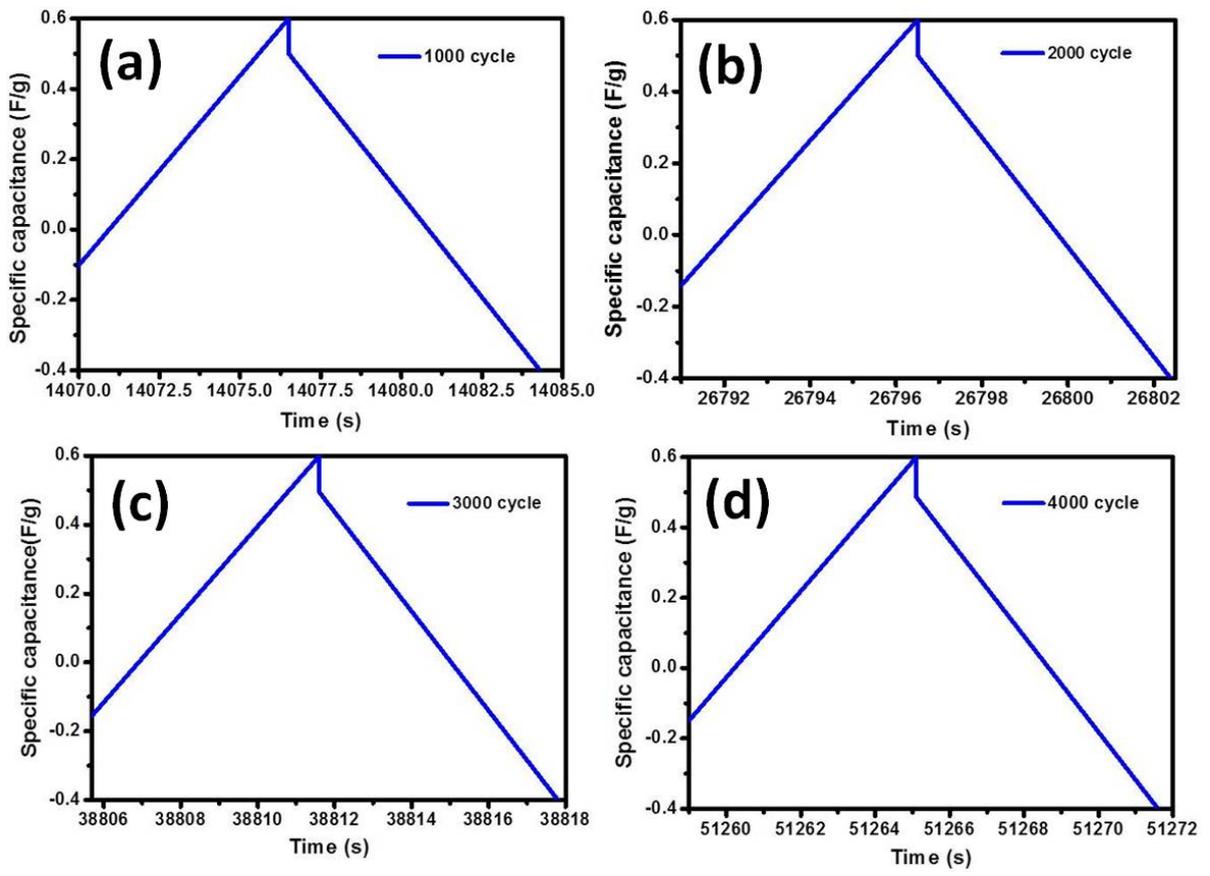

**Figure 4**